\documentclass[aps,twocolumn,showpacs,amsmath,amssymb,pra,superscriptaddress,floatfix,longbibliography]{revtex4-1}
\usepackage{braket}
\usepackage{amsmath}
\usepackage{amssymb}
\usepackage{mathtools}
\usepackage{graphicx}% Include figure files
\usepackage{dcolumn}% Align table columns on decimal point
\usepackage{bm}% bold math
\usepackage{multirow}
\usepackage{leftidx}
\usepackage{color}
\usepackage{float}

\usepackage{amsfonts}
\usepackage{eufrak}
\usepackage[german,english]{babel}

\begin{document}
\title{Simulations of Majorana spin flips in an antihydrogen trap}
\author{M.~A.~Alarc\'on}
\affiliation{Departamento de F\'isica, Universidad de Los Andes, Bogot\'a, Colombia}
\author{C.~J.~Riggert}
\affiliation{Homer L. Dodge Department of Physics and Astronomy, The University of Oklahoma,
Norman, Oklahoma 73019, USA}
\author{F.~Robicheaux}
\affiliation{Department of Physics, Purdue University, West Lafayette,
Indiana 47907, USA}
\email{robichf@purdue.edu}

\date{\today}

\begin{abstract}
The properties of antihydrogen ($\bar{\rm H}$) have, thus far, been probed at magnetic fields
of $\sim 1$~T. It may be fruitful to perform some of these measurements at magnetic
fields approaching 0~T. In this case,
there could occur zeros in the magnitude of the $B$-field. The number
and properties of the magnetic field zeros are investigated. For typical magnetic
field geometries in $\bar{\rm H}$ traps, the zeros will occur as two groups
of 5 closely spaced points instead of as a single point.
Except in special cases, results from calculations show that
these 10 zeros can be treated as independent sources of spin flip probability.
Although the behavior of Majorana spin flip near higher order zeros should
not be important in the $\bar{\rm H}$ traps, the probability for spin flip is
calculated for the case of a quadratic zero.
Finally, results are presented for a simple model of how
magnetic field zeros would affect the trapped population of $\bar{\rm H}$.
\end{abstract}

\maketitle
\section{Introduction}

More than 30 years ago, an effort was started to measure properties of
the antihydrogen ($\bar{\rm H}$) atom with the goal of comparing them with their matter
counterpart\cite{GRH}. Because the properties of H and $\bar{\rm H}$ should be exactly
the same by the CPT theorem, any difference would represent a fundamental
discovery\cite{BKR}. It is very difficult to generate any neutral antimatter atom or molecule
beyond $\bar{\rm H}$ which means it is fortunate that so many properties of
H are known to ultrahigh precision. In 2002, cold $\bar{\rm H}$
was experimentally formed at CERN\cite{AAB,GBO}.
In 2010, the ALPHA collaboration trapped $\bar{\rm H}$\cite{AAB2}
and within a year\cite{AAB3} demonstrated that the $\bar{\rm H}$ could
be held for an extensive time, sufficient for precision measurements.
To date, only the ALPHA collaboration has measured any property of the
$\bar{\rm H}$ atom although several groups are attempting to measure
various properties.
Examples of precision measurements  include the
hyperfine splitting of the $1S$ states\cite{AAB4,AAB6}, the charge of
the $\bar{\rm H}$\cite{AAB5,ABB}, the energy difference between
the $1S$ and $2S$ states\cite{ABB2,AAB7}, and the Lyman-$\alpha$ transition\cite{AAB8}.
Extensions of these measurements could lead to accurate determination of other
parameters. For example, a more accurate measurement of the Lyman-$\alpha$
transition would give the Lamb shift or the measurement of another narrow linewidth
transition (e.g. 2S-4S) would allow the determination of the antiproton radius
and the $\bar{\rm H}$ Rydberg constant.

The $\bar{\rm H}$ ground state has 4 non-degenerate levels in a magnetic field.
By convention these are labeled $1Sa,1Sb,1Sc,1Sd$ from lowest to highest
energy, see Fig.~1 of Ref.~\cite{AAB4}. The $1Sa,1Sb$ states have decreasing energy with increasing $B$ and,
thus, are high field seeking states which are expelled from a magnetic trap.
The $1Sc,1Sd$ states have increasing energy with increasing $B$ and can be
trapped. Above $\sim 0.1$~T, the states are effectively two pairs of states with
a magnetic moment approximately that of a free electron giving a slope of
$(dE/dB)/k_B \simeq \pm 2/3$~K/T. For small magnetic fields (less than $\sim 0.01$~T), the
states are more accurately represented as hyperfine eigenstates 
with an $F=0$ state 1420~MHz below
the $F=1$ states. For small $B$-field, the $F=1$ state is split to
$M_F=1,0,-1$ in increasing order of energy. The $M_F=0,-1$ are the states that adiabatically
connect to the trappable states.

Within the past few years, the ALPHA collaboration has successfully performed
several high precision measurements as enumerated in the first paragraph.
The trapping region is a tube
of length $\sim 25$~cm and radius $\sim 2$~cm. We will denote motion along the
axis to be axial motion represented by $z$ while the radial or angular motions
will be represented by $x,y$. Measurements in this trap have
taken place in magnetic fields of $\sim 1$~T which forces a comparison between
the measured $\bar{\rm H}$ transition frequencies and {\it calculated} frequencies
using the known properties of the positron and antiproton (e.g. masses, charges, and magnetic
dipole moments). If the magnetic field were smaller, then some of the terms in
the calculation of transition frequencies become irrelevant. As an example, the diamagnetic shift
of the $1S-2S$ frequency would be less than 0.4~Hz for $B=1$~mT\cite{RMR}. As another
example, the shift in energy due to the motional Stark effect was estimated to
be $\sim 300$~Hz in a 1~T field\cite{RMR} but would be much less than 1~Hz in a 1~mT
field since the shift is proportional to $B^2$. This suggests that the path to,
for example, $\sim 1-10$~Hz accuracy will be for the experiments to occur
at smaller magnetic field.

One of the difficulties of working at a smaller $B$-field is that it might accidentally
go to zero. In this case, the two trapped states, $M_F=0,-1$, could suffer a
Majorana spin flip if the $\bar{\rm H}$ passes too close to a $B$-field zero\cite{EM1,CZ1,LDL}.
The Majorana spin flip occurs because the body frame direction of the $B$-field
changes more rapidly than the precession frequency when passing near the
zero. For $\bar{\rm H}$, the situation is somewhat complicated because the
energies of the $F=1$ state are not exactly $-\mu B M_F$. However, for the
size of $B$ where the spin flip is possible, the linear dependence of the energy
on $B$ is good enough to obtain accurate spin flip cross sections. The main complicating
factor for traps like that in the ALPHA device is that there is more than one
zero and the zeros can be closely spaced, depending on the parameters, 5 zeros
can be separated by less than 1~mm. This special condition warrants an investigation
of the physics of spin-flip in this type of trap.

This paper is organized as follows. Section~\ref{SecZero} gives the possible
forms of the $B$-field near the zeros. Section~\ref{SecCrs} gives analytic
expressions for the spin flip cross section and rate which are accurate when
the zeros are separated. Section~\ref{SecComp} contains a comparison between
the analytic approximation to the cross section and a fully numerical result;
conditions are given for when the analytic approximation is accurate.
Section~\ref{SecMod} contains results for how the spin flip affects the
energy distribution of trapped $\bar{\rm H}$s. There is a short conclusions
section, Sec.~\ref{SecCon}. Section~\ref{SecApp} is a short appendix
that, for completeness, gives the derivation of the spin flip cross section for an isolated zero.

\section{Form of $\vec{B}$ near zeros}\label{SecZero}

For the Majorana spin flip process, the velocity of the $\bar{\rm H}$
and the variation of the B-field near $|\vec{B}|=0$
determines the spin flip probability. Away from the zero, the
$\bar{\rm H}$ magnetic moment adiabatically follows the magnetic
field direction. To get a sense of the relevant scales, the precession
of the positron spin is $\sim 30$~MHz at
1~mT. Since only $\bar{\rm H}$s
with kinetic energy less than $\sim 1/2$~K are trapped, their speed is
a few 10's~m/s. At 1~mT, the $\bar{\rm H}$ travels $\sim 1$~$\mu$m
during one precession period. The spatial variation of the magnetic
field near a zero is
$\sim 1$~T/m. Taking the change in $B$ during one precession period to
be $\sim 10\times$ smaller than $B$ suggests that only regions where the magnetic field
is less than $\sim 0.1$~mT are important for spin flip.

\begin{figure}
\resizebox{80mm}{!}{\includegraphics{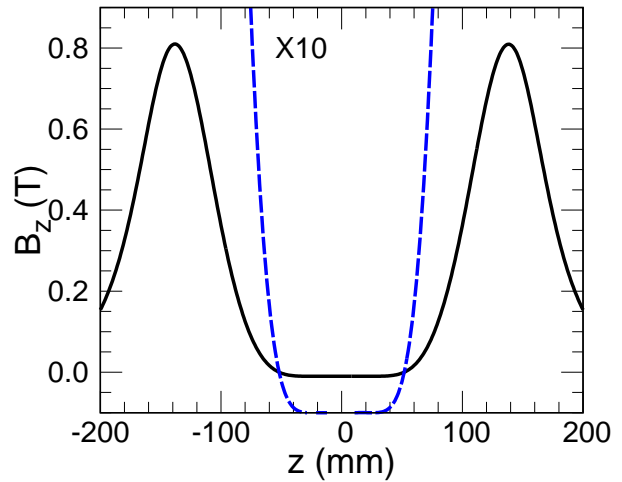}}
\caption{\label{FigBf}
$B_z$ on the trap axis as a function
of the axial position. $z=0$ corresponds to the trap center. This
magnetic field is the superposition from 5 mirror coils. The dashed (blue)
line is $10\times$ $B_z$.
}
\end{figure}

To obtain an idea of how many and where the magnetic zeros appear,
we numerically found the zeros for the magnetic field trap used in
Refs.~\cite{ABB2,AAB6,AAB7,AAB8} but shifted $B_z$ so that it was slightly negative,
$B_z=-0.01$~T, in the central region instead of $\sim 1$~T. A schematic
drawing of the trap is in any of these papers.
Mirror coils provide the axial, $z$, confinement while octupole coils provide
confinement in $x,y$.
The radius of the trap is approximately 22~mm.
The on axis $B_z$ is shown in Fig.~\ref{FigBf} where the trapping
region is between the $B_z$-maxima near $z=\pm 138$~mm. This magnetic
field is mainly generated with 5 mirror coils. The outer two coils give a
large positive $B_z$ on axis leading to the maxima.
The middle 3 mirror coils have opposite current (bucked) to the outer coils
giving a flattened B-field in the central region. This flattening is
desirable because it increases the precision of the spectroscopic
measurements and leads to a larger resonance region for the transitions
whose frequencies are shifted by $B$.
A nearly uniform $B$-field along $z$ sets the overall size of the on
axis field.
The octupole field is nearly zero on axis but plays a large role
off axis.

\begin{figure}
\resizebox{80mm}{!}{\includegraphics{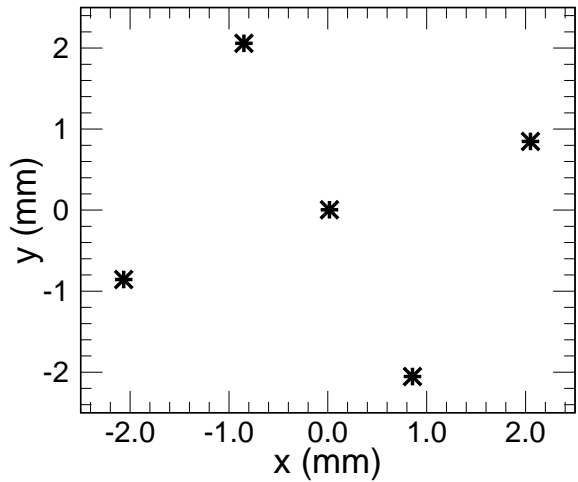}}
\caption{\label{FigZero}
Positions of the magnetic field zeros near $z=-52$~mm using the
full model of the ALPHA trap. The
central zero is actually $\simeq 16$~$\mu$m off the axis and
is at a slightly shifted axial position ($z=-51.84$~mm) compared
to the off axis zeros ($z=-51.96$~mm).
}
\end{figure}

The $10\times$ $B_z$ in Fig.~\ref{FigBf} makes clearer the two axial positions
where the $B$-field is near 0. There is a group of zeros near $z=-52$~mm and another
near $z=52$~mm.
Figure~\ref{FigZero} shows the $x,y$ positions of the zeros near
$z=-52$~mm. There is one zero nearly on axis and 4 that are at
nearly the same radius and separated by 90$^\circ$.
The off axis zeros near $z=52$~mm are rotated
by approximately 45$^\circ$ from those shown in Fig.~\ref{FigZero}.
Without the octupole field, there is only one zero and it is nearly on axis. The
combination of the octupole field and the radial component of the
magnetic field from the mirrors lead to the 4 off axis zeros.

The form of the magnetic field near the possible zeros is described in this
section for the case of an octupole field in $x,y$ plus a cylindrically symmetric
field that varies in $x,y,z$. This geometry is important for the antihydrogen
traps because both ALPHA and ATRAP have this magnetic field structure.
For an idealization of either apparatus, there is an octupole magnetic field
that increases with the radial distance from the center of the trap but whose magnitude has
no dependence on the axial coordinate. There will also be a cylindrically
symmetric magnetic field with a $z$-dependence on axis which is a
low power, e.g. $z^1$. Two or more mirror coils can generate axially
confining fields that are proportional to $z^2$ near the minimum or
higher power (e.g. $z^4$ or $z^6$) when using 5 mirror coils as in ALPHA.

This idealization of the magnetic field is accurate away from the trap walls
and in the region where the B-field is near its minimum value. The octupole
field seriously deviates from the idealization only near the end of the octupole
coils and near the walls ($r\simeq 22$~mm), but the zeros discussed below
are always within the axial central half of
the trap and far from the walls, see Fig.~\ref{FigZero} and the caption.
The B-fields from the mirrors only deviate from cylindrical symmetry
due to the leads or manufacturing imperfections. Thus, the deviations
from the ideal case should only lead to small linear terms near the zero
which will only slightly modify the variation of the magnetic field.
We compared our calculations of spin flip probability using the idealization to
those using a full model of all of the ALPHA coils and found only
negligible differences once the positions of the zeros were matched.

\subsection{Octupole magnetic field}\label{SecOct}

We will approximate the octupole field with the form
\begin{eqnarray}\label{EqBoct}
\vec{B}_o(x,y,z)&=&\frac{B_w}{r_w^3}(-x^3+3xy^2,-y^3+3yx^2,0)\nonumber\\
&=&\frac{B_w r^3}{r_w^3}[-\hat{e}_r\cos (4\phi )+\hat{e}_\phi \sin (4\phi )]
\end{eqnarray}
where $r_w$ is the radius of the trap wall,
$B_w$ is the magnitude of the octupole field at $r_w$,
$r^2=x^2+y^2$, $\hat{e}_r=\hat{e}_x\cos (\phi ) +\hat{e}_y\sin (\phi )$,
and $\hat{e}_\phi = -\hat{e}_x\sin (\phi ) + \hat{e}_y\cos (\phi )$.
This octupole field
has the property $|\vec{B}_o|=(r/r_w)^3 B_w$. The form of the octupole field
in Eq.~(\ref{EqBoct}) will
give zeros {\it that are rotated} from those shown in Fig.~\ref{FigZero}
but this rotation has no effect on the probability for a Majorana spin flip
when averaged over all possible trajectories from trapped  $\bar{\rm H}$s. We have chosen this
form, instead of that rotated to match Fig.~\ref{FigZero}, to simplify
the analysis below.

In all of the calculations below, we use $B_w/r_w^3=1.375\times 10^5$~T/m$^3$
which is a typical value used in ALPHA experiments.

\subsection{Octupole plus linear variation}\label{SecOpL}

The magnetic field which is cylindrically symmetric and has linear
spatial dependence is
\begin{eqnarray}\label{EqB1}
\vec{B}_c &=& \frac{B_1}{L}(x,y,-2(z-z_0))\nonumber\\
&=&\frac{B_1}{L}[r\hat{e}_r-2(z-z_0)\hat{e}_z]
\end{eqnarray}
where $-2B_1/L$ is the slope of $B_z$ on the axis at the zero
which is at $(0,0,z_0)$.

The zeros of the total magnetic field are the
positions where all components of $\vec{B}_o+\vec{B}_c$
are zero. Since the octupole field has no $z$-component, all of
the zeros are where $B_{c,z}=0$. This means all of the
zeros have $z=z_0$. From the caption of Fig.~\ref{FigZero},
this is a good approximation to the actual $B$-field where the differences 
in the $z$-position of the zeros are less than $\simeq 0.1$~mm.
The off axis zeros can be found by setting the coefficient of
$\hat{e}_\phi$ and $\hat{e}_r$ separately equal to 0.
Since the cylindrically symmetric field, $\vec{B}_c$, does
not have a $\hat{e}_\phi$ component, this conditions sets the angles of the
zeros from $\sin (4\phi_0 )=0$: $\phi_0 = n \pi /4$ with
$n=0, 1, 2, ..., 7$. Lastly, the coefficient of the $\hat{e}_r$
gives
\begin{equation}
-\frac{B_w}{r_w^3}\cos (4\phi_0 ) r_0^3 +\frac{B_1}{L}r_0 = 0
\end{equation}
which can only be zero if $\cos (4\phi_0 ) >0$ when $B_1/L>0$. This means only
4 of the angles allowed by the $\hat{e}_\phi$ condition give
zeros for the $\hat{e}_r$ condition: $\phi_0 = n \pi /2$ with
$n = 0, 1, 2, 3$.
If $B_1/L <0$, then the allowed angles are rotated by 45$^\circ$ from the
values for $B_1/L>0$. This feature  matches that in the actual
$B$-field.
At each of these angles, the radius is the same
value
\begin{equation}\label{EqRad}
r_0=\sqrt{\frac{B_1 r_w^3}{B_w L}}.
\end{equation}
These zeros have the same properties as those from the actual field:
the off axis zeros have nearly the same radius and are separated
by 90$^\circ$. As described in Sec.~\ref{SecOct},
 the angles do not match Fig.~\ref{FigZero} because of the choice of orientation
 of the octupole B-field
(chosen for simplicity of the resulting analysis).

To give an idea of sizes, the off axis zeros in Fig.~\ref{FigZero}
are at $r_0 \simeq 2.225\times 10^{-3}$~m. Using
$B_w/r_w^3=1.375\times 10^5$~T/m$^3$ from the previous
section gives $B_1/L = 0.681$~T/m.

For the Majorana spin flip, the variation of $\vec{B}$
in the neighborhood
of a zero is important. The spatially linear variation $(x,y,z)=(x_0+\delta x,y_0+\delta y,z_0+\delta z)$
has the form
\begin{equation}\label{EqBfL1}
\vec{B}= \frac{B_1}{L}(\delta x,\delta y,-2\delta z) + O(\delta ^3)
\end{equation}
for the on axis zero where $O(\delta^3)$ indicates the correction
is cubic in the position change. For the off axis zero at $\phi_0 = 0$ (i.e. on the
$x$-axis), a Taylor series expansion gives
\begin{equation}\label{EqBfL2}
\vec{B}= -\frac{2B_1}{L}(\delta x, -2\delta y, \delta z) + O(\delta^2)
\end{equation}
which has the same form as for the central zero except with a rotated
coordinate system and double the slope. All of the off axis zeros have
these properties: same form, double the slope of the central zero,
and rotated coordinate system.

An important question is whether the zeros will give independent spin
flip probabilities for most trajectories or whether the nearness of other
zeros will affect the spin flip process. As will be shown in Sec.~\ref{SecComp},
the zeros give independent contributions to the cross section as long
as the separation is larger than (approximately) the square root of the
spin flip cross section. This situation always occurs at large $B_1/L$
since the separation increases with increasing $B_1/L$
while the cross section decreases. The simulations show where
this approximation gives good results and where it is poor.

\subsection{Octupole plus quadratic variation}

The magnetic field which is cylindrically symmetric and has quadratic
spatial dependence is
\begin{eqnarray}\label{EqBf2}
\vec{B}_c&=& (0,0,B_0)+\frac{B_2}{L^2}(-xz,-yz,z^2 - \frac{1}{2}r^2)\nonumber\\
&=&\left(B_0 +\frac{B_2}{L^2}[z^2- \frac{1}{2}r^2]\right)\hat{e}_z-\frac{B_2}{L^2}rz\hat{e}_r
\end{eqnarray}
where the on axis minimum of $B_z$ is $B_0$ at $z=0$ and $B_2/L^2$ is
half the curvature of the B-field on axis. When only the outer coils make
the magnetic trap, the $B_2/L^2\sim 15$~T/m$^2$. When the field
is flattened as in Fig.~\ref{FigBf}, there can be inadvertent minima 
when attempting to obtain a flattened $B$-field with
$B_2/L^2\sim 1$~T/m$^2$ or somewhat smaller.

There can only be zeros on axis if $B_0<0$ and they are at
$z_0=\pm \sqrt{-B_0L^2/B_2}$ and $r_0=0$. The off axis case is
a bit more complicated. The $\hat{e}_\phi$ condition still gives
$\sin (4\phi_0 ) = 0$: $\phi_0 = n\pi /4$ with $n = 0,1,2, ...,7$.
The $\hat{e}_r$ condition gives
\begin{equation}
-\frac{B_2}{L^2}r_0z_0-\frac{B_w}{r_w^3}r_0^3\cos (4\phi_0) =0
\end{equation}
which restricts $\phi_0 = n\pi /2$ with $n=0,1,2,3$ if
$z_0<0$ and $\phi_0 = n\pi /4$ with $n=1,3,5,7$ if
$z_0>0$. These relations explain why the off axis zeros were rotated
by 45$^\circ$ in the actual magnetic field associated with Fig.~\ref{FigBf}.
This also gives the relationship $r^2_0={\cal L}|z_0|$ with
${\cal L}=(r_w^3/B_w)(B_2/L^2)$. The condition from $\hat{e}_z$
gives
\begin{equation}
z^2_0-\frac{1}{2}{\cal L}|z_0| - z^2_{0,a} = 0
\end{equation}
where $z^2_{0,a}=-B_0L^2/B_2$ is from the on axis zero.
There is a small range of cases where there are zeros with positive
$B_0$ but it is less than $\sim 10^{-8}$~T and, therefore, experimentally irrelevant.
For $B_0\leq 0$,
\begin{equation}
|z_0|=\frac{\cal L}{4}+\sqrt{({\cal L}/4)^2+z^2_{0,a}}\quad {\rm and}\quad
r_0=\sqrt{{\cal L}|z_0|}
\end{equation}
where $z^2_{0,a} \geq 0$.

It is worth considering the sizes of various terms using $B_2/L^2\sim 10$~T/m$^2$
and $B_w/r_w^3\sim 10^5$~T/m$^3$. The case $B_0=0$ is simple giving
$|z_0|=(B_2/L^2)/(2 B_w/r_w^3)$ and $r_0=|z_0|/\sqrt{2}$. This gives
a $z$-separation of $\sim 0.1$~mm and a similar size for $r_0$ implying
that $B_0=0$ might lead to interesting results. However, even relatively
small $B_0$ leads to the approximation in Sec.~\ref{SecOpL} working
well. For example, $B_0= -1$~mT and $B_2/L^2 = 10$~T/m$^2$
gives a separation in z of $20$~mm while changing to $B_0=0.1$~mT
gives a separation of 6.3~mm.
Therefore, only the $B_0=0$ case is probably of interest.

\subsection{Octupole plus quartic variation}

For the flatter potentials (like that pictured in Fig.~\ref{FigBf}), the
zeros become like the case of well separated linear zeros, Sec.~\ref{SecOpL}.
For example, in Fig.~\ref{FigBf}, the zeros
for the case $B_z(0,0,0) = -10$~mT gave a separation of $\sim 104$~mm
while $-1$~mT gave a separation of $\sim 80$~mm. It is likely that
imperfections in the magnetic field will mean this case will never be
experimentally interesting.

\section{Flip cross section and rate: linear zero approximation}\label{SecCrs}

As a baseline, the Majorana spin flip probability will be calculated for a
single zero. Since all of the zeros, Eqs.~(\ref{EqBfL1},\ref{EqBfL2}),
have a linear approximation of the form Eq.~(\ref{EqB1})
(except rotated), we only discuss the spin flip for that case. The time
dependent magnetic field at the atom is determined by the motion of the
atom which is assumed to be a straight line at constant speed. To simplify
the analysis, the origin of the coordinate system is at the zero of the
magnetic field. The position of the $\bar{\rm H}$
is given by
\begin{equation}\label{EqPos}
\vec{r}(t) =\vec{b} + \vec{v}t
\end{equation}
where $b$ is the impact parameter, $v$ is the speed, and we define the time
of closest approach as $t=0$ which means $\vec{b}\cdot\vec{v}=0$.
Since the position linearly depends on time and the magnetic field
linearly depends on the position, the magnetic field linearly depends
on time. For this case, Landau-Zener type theories can be used to
analytically obtain the transition probability between different states\cite{SLC}.
From the probability as a function of $\vec{b}$ and $\vec{v}$,
cross sections for particular transitions have been obtained
before\cite{CZ1,LDL,SLC}. For completeness, the derivation of the flip probability
is given in the appendix, Sec.~\ref{SecApp}.

For $\bar{\rm H}$, the upper two energy levels of the $F=1$ state
are the only ones that are trapped in the magnetic field. The upper
level is $M_F=-1$ and the next level is $M_F=0$. The cross section
for various flip processes is calculated from the transition probability
which is a function of $\vec{b}$ and $\vec{v}$ for a given $\bar{\rm H}$
speed, $v$. The derivation of the cross section for one linear zero
is given in the appendix, Sec.~\ref{SecApp},
\begin{eqnarray}\label{EqCr1}
\sigma_{1\leftarrow -1} &=&\sigma (v)\qquad {\rm and}\qquad
\sigma_{0\leftarrow -1} = 2\sigma (v)\nonumber\\
\sigma_{-1\leftarrow 0} &=&2\sigma (v)\qquad {\rm and}\qquad
\sigma_{1\leftarrow 0} = 2\sigma (v)
\end{eqnarray}
where $\sigma (v)=\hbar v /(\mu B_1/L)$ with $\mu = 9.28\times 10^{-24}$~J/T~$\simeq k_B 2/3$~K/T,
the magnetic moment of the positron.
This is an interesting result in that the flip rate, $v\sigma$,
is proportional to the kinetic energy of the $\bar{\rm H}$.
Thus, the atoms that are lost will tend to be the hottest.

For the case of the octupole plus linear variation in $z$, there were
5 zeros. The 4 off axis zeros had twice the slope as on the central
axis. If all 5 zeros give an {\it independent} contribution to the Majorana
flip cross section, the total cross section for the group of 5 will be
\begin{eqnarray}\label{EqCrsSec5}
\sigma_{1\leftarrow -1} &=&3\sigma (v)\qquad {\rm and}\qquad
\sigma_{0\leftarrow -1} = 6\sigma (v)\nonumber\\
\sigma_{-1\leftarrow 0} &=&6\sigma (v)\qquad {\rm and}\qquad
\sigma_{1\leftarrow 0} = 6\sigma (v)
\end{eqnarray}
where $\sigma (v)=\hbar v /(\mu B_1/L)$.
For the actual traps, there are two groups of 5 zeros implying
the total flip cross sections are double these results.

\section{Comparison to numerical}\label{SecComp}

In this section, the spin flip cross section from the numerical solution of the time
dependent Schr\"odinger equation is presented.

The time dependent Schr\"odinger equation was solved using the
Crank-Nicolson method\cite{CN1}:
\begin{equation}
\vec{\psi}(t+\delta t) =\frac{1-iH\delta t/(2\hbar )}{1+iH\delta t/(2\hbar )}\vec{\psi }(t)
\end{equation}
where the Hamiltonian, $H$ in Eq.~(\ref{EqHam}), is evaluated
at time $t+\delta t/2$. For spin-1, the Hamiltonian is a $3\times 3$ matrix
so the solution of this matrix equation is relatively fast.

The somewhat tricky aspect of obtaining the cross section is to determine
the fraction of population where $M_F$ has changed. The eigenstates
when $\vec{B}=B(\sin\theta\cos\phi , \sin\theta\sin\phi , \cos\theta )$ are
\begin{equation}\label{EqPsi}
\vec{\psi}_{\pm 1}=
\begin{pmatrix}
\frac{1}{2}(1\pm\cos\theta )e^{-i\phi}\\
\frac{1}{\sqrt{2}}\sin\theta\\
\frac{1}{2}(1\mp\cos\theta )e^{i\phi}
\end{pmatrix}
\quad
\vec{\psi}_0 =
\begin{pmatrix}
-\frac{1}{\sqrt{2}}\sin\theta e^{-i\phi }\\
\cos\theta\\
\frac{1}{\sqrt{2}}\sin\theta e^{i\phi }
\end{pmatrix}
\end{equation}
When solving the time dependent Schr\"odinger equation, the
magnetic field will start out in one direction and finish in
another. We used these equations to start the wave function
at the initial time and to project onto the final states.
In the calculations, we started the time propagation so that
the $\bar{\rm H}$ is far enough from the zeros that initially the
state adiabatically follows the changing direction of $\vec{B}$
and stopped the propagation when this condition was again
satisfied.

We found that using Eq.~(\ref{EqPos}) did not give results
that converged well with the starting and final time. The
problem is that starting with Eq.~(\ref{EqPos}) does not
adequately account for the slight difference between the adiabatic
and actual wave function unless the magnetic field is very large.
This causes the calculations to be quite slow because then the
wave function needs to be propagated for longer times {\it and}
the time steps need to be smaller to account for the larger
energy splittings.
We found that a $\vec{r}(t)$
where the velocity smoothly turned on from
0 to $\vec{v}$ and then smoothly turned back to 0 allowed for
accurate calculation of spin flip probabilities with relatively little
numerical effort. We used
\begin{equation}
\frac{d\vec{v}_{t}(t)}{dt} = \frac{\vec{v}}{\tau\sqrt{\pi }}\left[ e^{-(t-t_i)^2/\tau^2}
-e^{-(t-t_f)^2/\tau^2}\right]
\end{equation}
where $\tau$ is the duration of the turn-on
and $d\vec{r}(t)/dt=\vec{v}_{t}(t)$ for the time dependent position.
We then solved the time dependent Schr\"odinger equation
from $t_i-6\tau$ to $t_f + 6\tau$. As long as $\tau$ was much larger
than $\hbar /(\mu B)$ with $B$ evaluated at the starting and final time,
then the convergence was much faster with respect to $t_i,t_f$.

For the calculation of the cross section, we used a Monte Carlo
sampling of the $\hat{v}$ and $\vec{b}$. The random parameters
were chosen as: $b^2$ randomly chosen with a flat distribution
between 0 and $b^2_{max}$,
$\hat{v}$ randomly chosen with a flat distribution on the surface of a
unit sphere, and $\hat{b}$ randomly chosen from a flat distribution on
the great circle defined by $\hat{v}\cdot\hat{b}=0$. The cross section
for a transition is the average probability for that transition times $\pi b^2_{max}$.
We checked for convergence with respect to $b_{max}$ and the
number of trajectories. The $b_{max}$ can be estimated from
Eq.~(\ref{EqLZ}) by setting $\Gamma > 16/\pi$ for all angles and adding this to
the $r_0$ of the off-axis zeros.

As a test of the program, we solved for the spin flip cross section for the pure linear
$B$-field, Eq.~(\ref{EqB1}). We found that the cross section only differed
from the analytic value, Eq.~(\ref{EqCr1}), due to statistical sampling.

\subsection{Octupole plus linear variation}

In this section are the numerical results for
the case of the octupole plus linearly varying $\vec{B}_c$ described in Sec.~\ref{SecOpL}.
For this case, we compared the cross section for independent contribution
from the 5 zeros, Eq. (\ref{EqCrsSec5}), to that from a numerical calculation.
In the numerical calculation, we ran approximately 200,000 trajectories to
obtain adequate statistics for the Monte Carlo cross section.

The case shown in Figs.~\ref{FigBf} and \ref{FigZero} has a linear parameter
$B_1/L=0.681$~T/m. Using these parameters and $v=50$~m/s, we found
the Monte Carlo result to be the same as Eq. (\ref{EqCrsSec5}) within
the statistical uncertainty. For this case, $\sigma (v) = 8.33\times 10^{-10}$~m$^2$.

\begin{figure}
\resizebox{80mm}{!}{\includegraphics{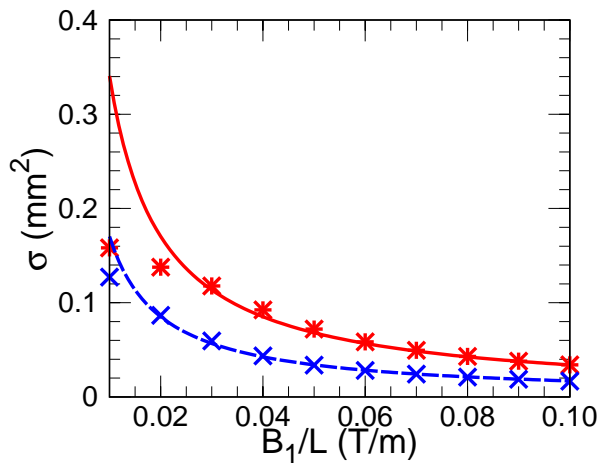}}
\caption{\label{FigFlip1}
The numerically calculated flip cross sections, $|\Delta M_F|=1$ (red $*$)
and $|\Delta M_F|=2$ (blue $\times$), compared to the separated zero
approximation, Eq.~(\ref{EqCrsSec5}), solid red line and blue dashed line.
The statistical uncertainty is less than the size of the symbols.
}
\end{figure}

To understand when to expect the separated zero approximation, Eq.~(\ref{EqCrsSec5}),
to fail, we plot the numerically calculated cross sections versus $B_1/L$
in Fig.~\ref{FigFlip1}. As in the approximation in Eq.~(\ref{EqCrsSec5}),
we found that the cross sections for $|\Delta M_F|=1$ were all the same
and those for $|\Delta M_F|=2$ were all the same. We also show the
results for the separated zero approximation, Eq.~(\ref{EqCrsSec5}),
as a comparison.
As can be seen, there starts to be noticeable differences when
$B_1/L <0.03$ for the $|\Delta M_F|=1$ case. At the lowest $B_1/L$
calculated (0.01~T/m), the numerical result is more than a factor of 2 smaller
than the separated zero approximation. The  $|\Delta M_F|=2$ is better
matched by the approximation with substantial difference only for the
smallest $B_1/L$. It seems reasonable that the separated zero approximation
will break down when the flip cross section equals $\pi r_0^2$ where
the $r_0$ is the radius of the off axis zeros, Eq.~(\ref{EqRad}). For
the parameters in this section, this condition gives $B_1/L\sim 0.022$~T/m
for  the $|\Delta M_F|=1$ case and $B_1/L\sim 0.015$~T/m for the
$|\Delta M_F|=2$ case. These values are reasonably close to where
the differences begin to appear in Fig.~\ref{FigFlip1}.
For the $|\Delta M_F|=1$ case, this condition
is $B_1/L=\sqrt{6\hbar vB_w/(\mu r_w^3)}$ and is $\sqrt{2}$ smaller
for the  $|\Delta M_F|=2$ case. Experimental control
of magnetic fields at this level is possible\cite{AAB6,AAB7}.

\subsection{Octupole plus quadratic variation}

For this section, we will consider the cases where 1~T/m$^2 \leq B_2/L^2\leq 10$~T/m$^2$
which is a reasonable range for the $\bar{\rm H}$ traps.
The case of an octupole field plus quadratic cylindrical field, Eq.~(\ref{EqBf2}),
has 3 situations worth considering for these parameters: $B_0>0.025$~mT,
$B_0 = 0$~T, and $B_0 < -0.05$~mT.

The case $B_0>0.025$~mT is the simplest. The cross section is, within numerical
errors, consistent with 0. The spin can adiabatically follow the changing direction
of $\vec{B}$ for this case. We did not test how small can $B_0$ be before the
cross section is non-negligible since $0.025$~mT is already below the accuracy
for the experimental values of trap parameters.

\begin{figure}
\resizebox{80mm}{!}{\includegraphics{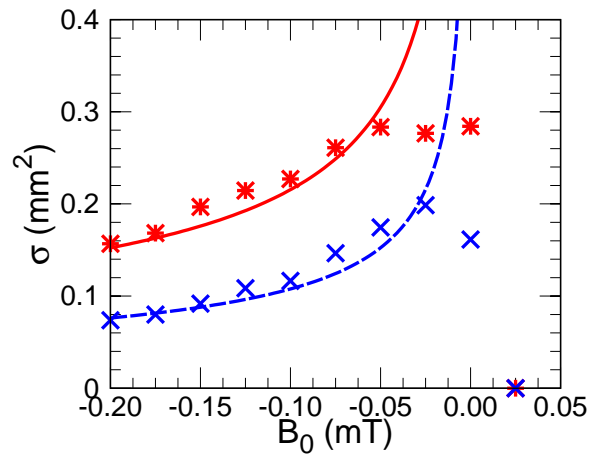}}
\caption{\label{FigFlip2}
The numerically calculated flip cross sections, $|\Delta M_F|=1$ (red $*$)
and $|\Delta M_F|=2$ (blue $\times$), compared to the separated zero
approximation, Eq.~(\ref{EqCrsSec10}), solid red line and blue dashed line.
The statistical uncertainty is less than the size of the symbols
}
\end{figure}

The next simplest case is $B_0 < -0.05$~mT. In this situation, there are two
groups of 5 zeros near $|z_0|\simeq \sqrt{-B_0L^2/B_2}>1.6$~mm.
In this case, the $B$-field in the neighborhood of the zeros is approximately
that of the linear variation case with $B_1/L\simeq z_{0,a}B_2/L^2=\sqrt{-B_0B_2/L^2}$.
In this limit of $B_0$, the zeros are relatively separated so the approximation
in Eq.~(\ref{EqCrsSec5}) can be used.
For this case, the 10 zeros sum to give cross sections
\begin{eqnarray}\label{EqCrsSec10}
\sigma_{1\leftarrow -1} &=&6\sigma (v)\qquad {\rm and}\qquad
\sigma_{0\leftarrow -1} = 12\sigma (v)\nonumber\\
\sigma_{-1\leftarrow 0} &=&12\sigma (v)\qquad {\rm and}\qquad
\sigma_{1\leftarrow 0} = 12\sigma (v)
\end{eqnarray}
where $\sigma (v)=\hbar v /(\mu \sqrt{-B_0B_2/L^2})$. The numerical
results are compared to this approximation in Fig.~\ref{FigFlip2}.
As with the comparison in Fig.~\ref{FigFlip1} for the linear zero,
the agreement between the separated zero approximation and the numerical
result is good until the separation of the zeros is comparable to the
square root of the cross section.

The last important range is $B_0\sim 0$. For this value of $B_0$, the cross section
is finite instead of diverging as in Eq.~(\ref{EqCrsSec10}). The approximate size is that for the independent
zeros where the square root of the cross section is comparable to the separation
of the zeros as with the case in Fig.~\ref{FigFlip1}.

\section{Antihydrogen loss model}\label{SecMod}

This section contains results for a simple model calculation for the loss of $\bar{\rm H}$s from
a trap for the conditions of Figs.~\ref{FigBf} and \ref{FigZero}. For this case
there are 10 well separated zeros and the cross section for the various flip
processes are twice the values in Eq.~(\ref{EqCrsSec5}) with $B_1/L=0.681$~T/m.

In the ALPHA experiment, the volume of the trap is  $\sim 100$~cm$^3$ and
they have demonstrated trapping of $\sim 100$ atoms\cite{AAB9,AAB7,AAB8}. Thus, for absolute numbers
we will take the $\bar{\rm H}$ density to be 1~cm$^{-3}$. Because the trap depth
is only $\sim\frac{1}{2} k_B$K and the $\bar{\rm H}$s are formed at much higher temperatures,
the distribution of atoms is approximately a flat distribution in velocity space within
a sphere corresponding to a kinetic energy of $\simeq \frac{1}{2} k_B$K leading
to a normalized distribution with respect to speed of
$P(v) = 3 v^2/v_{\rm max}^3$. This distribution gives good
agreement with measurements\cite{AAB10,AAB8}. From these parameters,
we can estimate the rate for a spin flip process with cross section
$\sigma = C \sigma (v)$ with $\sigma (v)=\hbar v/(\mu B_1/L)$ to be
\begin{equation}
\Gamma =\rho\frac{3}{v^3_{\rm max}}\int_0^{v_{\rm max}} C v \sigma (v) v^2 dv
=\frac{3}{5}\rho v_{\rm max}C\sigma (v_{\rm max})
\end{equation}
where $\rho$ is the number density of $\bar{\rm H}$s and $v_{\rm max}\simeq 91$~m/s
for $KE_{max}=\frac{1}{2} k_B$K. Using these numbers, the $|\Delta M_F|=2$ rate (uses $C=6$) is
$\Gamma\simeq 0.83$~s$^{-1}$ and the $|\Delta M_F|=1$ rate (uses $C=12$)
is twice this value.
In the ALPHA experiment, the two trapped states are formed with equal probability
so the loss rate is the average of these, $\Gamma \simeq 1.2$~s$^{-1}$.
This suggests that the zeros can not be present for more than a couple 10's of seconds
before a substantial
fraction of the $\bar{\rm H}$s are lost.

\begin{figure}
\resizebox{80mm}{!}{\includegraphics{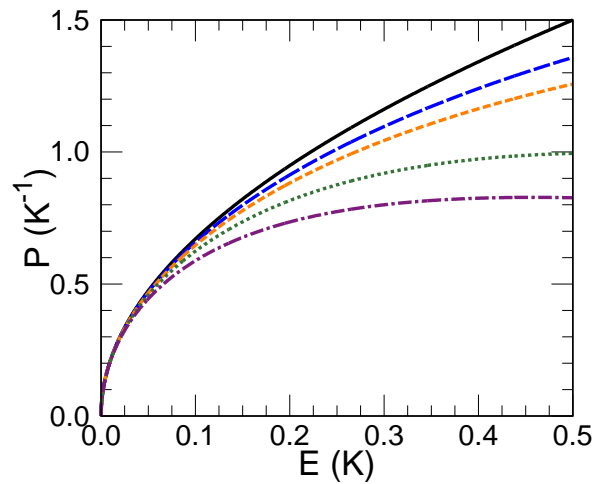}}
\caption{\label{FigRate}
The energy distribution of  $\bar{\rm H}$s after different amounts of interaction
with $B$-field zeros. Both the $M_F=0$ and $-1$ states start with the distribution
given by the black solid line. At the time when 10\%
of the $\bar{\rm H}$s have been lost, the $M_F=-1$ distribution is the dashed (blue)
line and the $M_F=0$ distribution is the short dashed (orange) line.
At the time when 30\%
of the $\bar{\rm H}$s have been lost, the $M_F=-1$ distribution is the dotted (green)
line and the $M_F=0$ distribution is the dot-dashed (purple) line.
}
\end{figure}

An important question is how the populations evolve if the zeros are present for
a substantial amount of time. This is not completely trivial because the higher
energy $\bar{\rm H}$s are preferentially lost {\it and} because the zeros mix the two
trapped states as well as leading to loss. We solved the coupled rate equations
at each velocity for the flat velocity distribution described above with
$E_{\rm max}=\frac{1}{2}k_B$K.
The $\bar{\rm H}$s were started
with a flat distribution in velocity and equal probability in the $M_F=0$ and $-1$ states:
the energy distributions are ${ P}_{-1}(E)={ P}_0(E)=(3/4)E^{1/2}/E_{\rm max}^{3/2}$.
We then evolved the distributions using twice the rates from
Eq.~(\ref{EqCrsSec5}) with $B_1/L=0.681$~T/m.
The results are shown in Fig.~\ref{FigRate} for the initial distribution and when
10\%
and when 30\%
of the $\bar{\rm H}$s have been lost. As can be seen, the higher energy $\bar{\rm H}$s 
are preferentially lost. Also, the distribution goes from equal population in $M_F=0$
and $-1$ to a larger fraction of $M_F=-1$. This is because the loss rate from
$M_F=0$ is larger.

There are two major limitations of the model. The first is that the higher energy $\bar{\rm H}$s 
are in, effectively, a spatially larger trap. Although the trapping potential in
Refs.~\cite{ABB2,AAB6,AAB7,AAB8}
are relatively flat, it is not an infinite square well. This means that the higher energy
$\bar{\rm H}$s  will pass by the zeros less often than in the simple model calculation.
The second is that the mixing between the different degrees of freedom takes some time
which can lead to depletion of certain types of trajectories. Again, this will lead to
a somewhat smaller loss rate for the regions of phase space that does not mix quickly.
Reference~\cite{ZFZ} found that the higher energy trajectories tended to mix more
quickly which might somewhat counteract the effect of the somewhat larger trap
volume at higher energy.

\section{Conclusions}\label{SecCon}

A description was given of the type of $B$-field zeros expected in $\bar{\rm H}$
traps. The octupole magnetic field that gives trapping in the radial direction leads
to the case where the zeros will typically be in two groups of 5 zeros with the two
groups having a large axial separation. The spacing of the zeros within a group of
5 is proportional to the square root of the slope of the $B$-field on axis and inversely
proportional to the octupole strength. The cross section for the Majorana spin flip
is proportional to the speed of the $\bar{\rm H}$ and inversely proportional to
the slope of the magnetic field on axis at the $B$-field zero. Interestingly, for typical octupole
trapping fields, the cross section is independent of the octupole field strength.
Numerical calculation of the spin flip cross section were performed and were compared
to analytic expressions for the spin flip cross section. The analytic cross sections are accurate
as long as the square of the separation of the zeros is larger than the spin flip
cross sections.

The evolution of the trap population was calculated for a simple model. In this model,
the $\bar{\rm H}$s  have a flat velocity distribution up to a maximum energy; this
maximum energy is the trap depth. This simple model showed that the higher energy
$\bar{\rm H}$s  are preferentially lost and that an equal distribution of $M_F=-1$
and $0$ states becomes somewhat biased to $M_F=-1$.

The results presented above may be useful in designing a strategy for performing
experiments on $\bar{\rm H}$ with small $B$-fields. For example, since the loss rate 
for an individual $\bar{\rm H}$ for typical
parameters is $\sim 10^{-2}$~s$^{-1}$ and that the loss rate is highest just after
the appearance of the zeros, an experiment might slowly lower the uniform $B$-field
until the Majorana spin flips start occurring. At that point, the $B$-field can be increased
to the point that the flips stop. As long as this manipulation occurs on a time scale less
than a couple 10's of seconds, there will not be a substantial loss of $\bar{\rm H}$s.
Finally, it is possible to use these zeros to diagnose properties of the magnetic field
that might be useful in experiments. For example, in a measurement of the effect
of gravity on $\bar{\rm H}$s, it is important to not have a spatial gradient in the
vertical direction which can mimic the force from gravity. By comparing the expected
and measured positions where the spin flips occur, the size of a spatial gradient in the
vertical direction can be diagnosed.

\section{Acknowledgment}
MAA was supported by the Purdue University SURF program,
CJR was supported by
the National Science Foundation under Award No. 1460899-PHY
REU program, and FR was supported by the
National Science Foundation under Award No. 1806380-PHY.

\section{Appendix}\label{SecApp}

This section gives the derivation of the spin flip cross section for completeness.

Start with the form for the magnetic field, Eq.~(\ref{EqB1}) with $z_0=0$, and the
position as a function of time, Eq.~(\ref{EqPos}). Define the magnitudes $v,b$ and
the angles $\alpha ,\beta $ where
\begin{eqnarray}
\vec{v}&=&v(\sin (\beta ),0,\cos (\beta ))\nonumber\\
\vec{b}&=&b(\cos (\alpha )\cos (\beta ),\sin (\alpha ),-\cos (\alpha )\sin (\beta )).
\end{eqnarray}
In this equation, advantage has been taken of the cylindrical symmetry of the magnetic
field to define the velocity vector to be in the $xz$ plane. The time dependent
magnetic field is then
\begin{eqnarray}
\vec{B}(t)&=&\frac{B_1b}{L}(\cos (\alpha )\cos (\beta ),\sin (\alpha ),2\cos (\alpha )\sin (\beta ))\nonumber\\
 &\null &+\frac{B_1v}{L}(\sin (\beta ),0,-2\cos (\beta )) t.
\end{eqnarray}
The coordinate system is now rotated so that the term multiplying $t$ is
purely in the $z$-direction and the constant part of $B_z$ is removed
by defining $t=0$ as the time of smallest $|\vec{B}|$:
\begin{eqnarray}
\vec{B}(t)&=&\frac{B_1b}{L}(\frac{2\cos (\alpha )}{\sqrt{1+3\cos^2(\beta )}},\sin (\alpha ),0)\nonumber\\
 &\null &+\frac{B_1v}{L}\sqrt{1+3\cos^2(\beta )}(0,0,1) t.
\end{eqnarray}
Lastly, rotate in the $xy$ plane so that $B_y=0$ to find
\begin{eqnarray}
\vec{B}(t)&=&\hat{e}_x\frac{B_1b}{L}\sqrt{\frac{4\cos^2(\alpha )}{1+3\cos^2(\beta )}+\sin^2 (\alpha )}\nonumber\\
 &\null &+\hat{e}_z\frac{B_1v}{L}\sqrt{1+3\cos^2(\beta )} t\nonumber\\
 &\equiv&(B_x,0,\dot{B}_zt)
\end{eqnarray}
which defines the size of the transverse magnetic field and the size
of the time derivative of the magnetic field along $z$. The Hamiltonian
for the spin system is defined as
\begin{equation}\label{EqHam}
H = -\frac{\mu}{s\hbar}\vec{B}(t)\cdot\vec{S}
\end{equation}
where $s=1$ for the $F=1$ case and would be $s=1/2$ for a spin-1/2 system.

We will first treat the more familiar spin-1/2 system because there is only
one spin flip possibility.
Using Landau-Zener formalism,\cite{EM1,CZ1,LDL,SLC} the spin flip probability for a spin-1/2 system
would be
\begin{equation}\label{EqLZ}
P_{-1/2\leftarrow 1/2}=e^{-2\pi\Gamma}\quad {\rm with}\quad
\Gamma = \frac{\mu B_x^2}{2\hbar\dot{B}_z}.
\end{equation}
To obtain the cross section for the spin flip, the probability needs to be averaged
over $\cos (\beta)$ and $\alpha$ and integrated over $2\pi b db$:
\begin{equation}
\sigma (v) = \frac{2\pi}{4\pi }\int_0^\infty \int_{-1}^1\int_0^{2\pi}
P_{-1/2\leftarrow 1/2}    d\alpha  d (\cos\beta ) b db
\end{equation}
Perform the integration with respect to $bdb$ first to obtain
\begin{equation}
\sigma (v) = \frac{vL\hbar}{4\pi\mu B_1}\int_{-1}^1\int_0^{2\pi}
\frac{(1+3s^2)^{3/2}}{4-3(1-s^2)\sin^2 (\alpha )}d\alpha ds
\end{equation}
where the transformation of variables $s=\cos\beta$ was used. Integrating over
$\alpha$ gives
\begin{equation}
\sigma (v)= \frac{vL\hbar}{4\pi\mu B_1}\int_{-1}^1 \pi (1 + 3s^2)ds=\frac{\hbar v}{\mu B_1/L}
\end{equation}

The case for spin 1 can be done analytically using a result from Ref.~\cite{SLC}.
The parameters for the Majorana spin flip can be converted to their parameters:
\begin{eqnarray}
b_1 &=& -b_3 = -\frac{\mu B_1 v}{L\hbar}\sqrt{1 + 3\cos^2\beta}\nonumber\\
g_1&=&g_2 = -\frac{\mu B b}{L\hbar\sqrt{2}}\left( \frac{\cos^2\alpha}{1+3\cos^2\beta}+\sin^2\alpha  \right)^{1/2}
\end{eqnarray}
and $b_2=0$. This leads to probabilities of the same form as for the spin 1/2 system
so that after integrating over impact parameter and averaging over $\alpha ,\cos\beta $,
the cross sections have the same form:
\begin{eqnarray}
\sigma_{1\leftarrow -1} &=&\sigma (v)\qquad {\rm and}\qquad
\sigma_{0\leftarrow -1} = 2\sigma (v)\nonumber\\
\sigma_{-1\leftarrow 0} &=&2\sigma (v)\qquad {\rm and}\qquad
\sigma_{1\leftarrow 0} = 2\sigma (v)
\end{eqnarray}
where we have only included transitions out of the trapped $M_F=0,-1$ states.

\bibliography{majorana}

\end{document}